\documentclass[aps,prm,twocolumn, groupedaddress]{revtex4-2}
\usepackage{graphicx}
\usepackage{upgreek}
\usepackage{float}

\begin{document}

\title{Epitaxial growth and magnetic phase transitions in non-centrosymmetric EuPdSi$_3$ thin films}

\author{Sebastian K\"olsch}
\email[Corresponding author. E-mail: ]{koelsch@physik.uni-frankfurt.de}
\affiliation{Thin Films and Nanostructures, Physical Institute, Goethe University Frankfurt, Max-von-Laue Strasse 1, Frankfurt am Main 60438, Germany}

\author{Alfons G. Schuck}
\affiliation{Slow Dynamics in Quantum Materials, Physical Institute, Goethe University Frankfurt, Max-von-Laue Strasse 1, Frankfurt am Main 60438, Germany}

\author{Olena Fedchenko}
\affiliation{Solid-state spectroscopy of electronically correlated materials, Physical Institute, Goethe University Frankfurt, Max-von-Laue Strasse 1, Frankfurt am Main 60438, Germany}

\author{Dmitry Vasilyev}
\author{Olena Tkach}
\affiliation{Group Magnetism, Institute of Physics, Johannes Gutenberg University, Staudingerweg 7, Mainz 55128, Germany}

\author{Sergeij Chernov}
\author{Christoph Schl\"uter}
\author{Andrii Gloskowski}
\affiliation{Photon Science, Deutsches Elektronen-Synchrotron DESY, Notkestrasse 85, 22607 Hamburg, Germany}

\author{Hans-Joachim Elmers}
\author{Gerd Sch\"onhense}
\affiliation{Group Magnetism, Institute of Physics, Johannes Gutenberg University, Staudingerweg 7, Mainz 55128, Germany}

\author{Jens Müller}
\affiliation{Slow Dynamics in Quantum Materials, Physical Institute, Goethe University Frankfurt, Max-von-Laue Strasse 1, Frankfurt am Main 60438, Germany}

\author{Michael Huth}
\affiliation{Thin Films and Nanostructures, Physical Institute, Goethe University Frankfurt, Max-von-Laue Strasse 1, Frankfurt am Main 60438, Germany}

\begin{abstract}
Non-centrosymmetric magnetic materials are a promising platform for stabilizing chiral spin textures, such as skyrmions and cycloidal magnetic states. This is particularly true in epitaxial thin film geometries, where strain and interface effects offer additional control. 
Herein, we report on the first epitaxial thin films of EuPdSi$_3$ grown by molecular beam epitaxy on MgO(001). 
X-ray diffraction confirms an epitaxial relationship of tetragonal EuPdSi$_3$ in the BaNiSn$_3$ structure with out-of-plane c-axis orientation and parallel in-plane a-axes. 
Hard x-ray photoelectron spectroscopy reveals a stable Eu valence of 2.0, yielding a large magnetic moment of approximately 7\,$\upmu_{\text{B}}$ per Eu atom in accordance with Hund’s rule. 
Owing to the non-centrosymmetric crystal structure, non-collinear magnetic states such as N\'{e}el-type skyrmions and cycloidal phases are allowed by symmetry. 
Electronic transport measurements reveal two magnetic phase transitions at 19\,K and 15\,K in zero applied field. 
Under magnetic fields applied along the crystallographic [100] and [001] directions, distinct temperature dependent magnetic phases emerge, demonstrating the sensitivity of the magnetic ground state to field orientation in epitaxial EuPdSi$_3$ thin films.
\end{abstract}

\maketitle
\section{Introduction}
Europium-based intermetallic compounds constitute a particularly intriguing class of materials at the intersection of correlated-electron physics, magnetism, and valence transitions or crossovers. 
Their experimental investigation in single crystalline bulk or thin film form, however, remains limited compared to other lanthanoids. 
This scarcity is largely caused by the intrinsic difficulties associated with Europium as a constituent element \cite{ramarao_on_2020}: its exceptionally high vapor pressure and strong chemical reactivity render controlled growth challenging. 
As a consequence, the majority of Eu-based intermetallics have so far been studied almost exclusively in the form of bulk polycrystals, while only recent achievements in this active research area yielded bulk single crystals of high quality, see e.\,g. \cite{onuki_divalent_2017,kliemt_strong_2022}.
Nevertheless, investigations of epitaxial thin films are lacking, leaving fundamental questions regarding dimensionality, strain effects, interface phenomena and the feasibility of integrating these materials into functional heterostructures unanswered.\\

However, a variety of interesting phenomena such as heavy-fermion like behavior, complex magnetic ordering or significant valence changes under external stimuli, e.\,g. temperature, magnetic field or pressure, appear in these compounds \cite{adams_effect_1991, onuki_divalent_2017}. 
Divalent (4f$^7$) and trivalent (4f$^6$) Eu states can be quite close in energy, as compared to most other lanthanoids, usually having a stable trivalent electronic configuration \cite{gschneidner_on_1968,melsen_calculation_1994}.
Yet, both valence states of Europium differ substantially in terms of their magnetic moment and ionic size, which manifests \textit{inter alia} in distinct lattice parameters, as can be observed in the lanthanoid series of isostructural compounds.
Owing to this notably strong interconnection between lattice and electronic degrees of freedom the influence of strain in epitaxial thin films is particularly interesting and represents another type of external control of the material properties, for example, of the magnetic ordering temperature in Eu-based intermetallics \cite{schuck_strain_2024}.

\newpage
Among the structurally related families of rare-earth intermetallics, compounds derived from the BaAl$_4$ structure type are the most common \cite{ramarao_on_2020}.
This family encompasses both centrosymmetric and non-centrosymmetric ternary variants, most prominently the tetragonal ThCr$_2$Si$_2$-type structure (I4/mmm) and the BaNiSn$_3$-type structure (I4mm), for a recent review see \cite{shatruk-thcr2si2-2019}. 
While the ThCr$_2$Si$_2$ structure has served as a paradigmatic platform for studying magnetism and electronic correlations in rare-earth based intermetallics \cite{lai_electronic_2022}, the BaNiSn$_3$ structure lacks a mirror plane perpendicular to the \textit{c}-axis, resulting in a broken inversion symmetry at the Ba site.\\

This subtle but decisive reduction in crystallographic symmetry has profound consequences for the electronic and magnetic ground states.
Accordingly, Dzyaloshinsky-Moriya-interactions (DMI) are allowed by symmetry, favoring a perpendicular alignment of magnetic moments.
Consequently, depending on the ratio of the coupling strength of the DMI and the strength of the RKKY exchange interaction between adjacent Eu-moments, long-wavelength modulations of the magnetic structure are to be expected.
An illustrative example can be found in EuPtSi$_3$, showing two magnetic phase transitions at 16\,K and 17\,K at zero applied field \cite{bauer_magnetic_2022}.
Furthermore, upon application of a magnetic field along different crystallographic directions metamagnetic behavior is observed, yielding cycloidal, conical or fan-like superstructures.
More interestingly, N\'{e}el-type skyrmions are symmetry-allowed for the BaNiSn$_3$-structure, but were not found in EuPtSi$_3$ \cite{simeth_resonant_2023}.
Although first T-H phase diagrams of the isostructural compound EuPdSi$_3$ are known \cite{yonehara_single_2020, nakashima_magnetic_2026}, no measurements of the actual magnetic structures are available, leaving the question regarding skyrmionic structures open.\\

In this work, we report about the first successful growth of epitaxial EuPdSi$_3$ thin films on MgO(100) substrates by molecular beam epitaxy (MBE) and present a comprehensive characterization of their structural and temperature dependent electronic properties. 
Reflection high-energy electron diffraction (RHEED) is employed to monitor the growth and crystalline order \textit{in-situ}, while Hard x-ray Photoelectron Spectroscopy (HAXPES) provides bulk-sensitive insights into the valence state of Eu and the electronic configuration.
Complementary electronic transport measurements reveal characteristic signatures of non-collinear magnetic structures in this non-centrosymmetric Eu-based intermetallic upon cooling.

\section{Experimental Procedure}
Epitaxial EuPdSi$_3$ thin films were grown using MBE in an ultra-high vacuum (UHV) chamber with a base pressure well below $1\times10^{-8}$\,Pa, dedicated to the preparation of rare-earth based intermetallics, see e.\,g¸. \cite{koelsch_clamping_2022, koelsch_epitaxial_2024}.
During growth, the pressure in the chamber remains lower than p\,=\,$5\times 10^{-7}$\,Pa, whereby the residual atmosphere consists mainly of hydrogen, which desorbs from the heated Eu material inside the effusion cell.\\

EuPdSi$_3$ crystallizes in the tetragonal BaNiSn$_3$ structure (space group 107) with lattice constants a\,=\,4.283\,\AA\, and c\,=\,9.897\,\AA\, \cite{nakashima_magnetic_2026}. 
Since no information of the lattice constants of EuPdSi$_3$ at elevated temperatures are available, the misfit between substrate and film was determined at room temperature to be $(a_{\text{EuPd}\text{Si}_3}-a_{\text{MgO}})/a_{\text{EuPd}\text{Si}_3}\,=\,1\,\%$.
Thus, the chosen substrate material has a small misfit, implying a cuboid-on-cube growth with an out-of-plane \textit{c}-axis orientation for the EuPdSi$_3$ thin film.
This was also observed for the growth of EuPd$_2$Si$_2$(001) on MgO(100), see \cite{koelsch_clamping_2022}.
Epi-polished (R$_a\,<$\,0.5\,nm) MgO substrates with $<$100$>$-out-of-plane orientation were obtained from Crystec GmbH.
Owing to the hygroscopy of MgO, the substrates were stored in a desiccator until insertion into the vacuum system.
Additionally, no chemical cleaning was performed, due to the highly probable formation of hydroxyl and carbonate groups on the surface \cite{febvrier_wet_2017,braun_situ_2020}.
Instead, the substrates were degassed in the loadlock chamber at 773\,K and additionally annealed at 1273\,K under UHV conditions in the growth chamber for at least one hour, to desorb contaminations from the surface.\\

To prevent oxidation of the grown thin films during investigations under atmospheric conditions, the samples were capped afterwards with an amorphous Si layer in the same chamber after cooling down to room temperature.
The typical Si capping layer thickness was approx. 3\,nm.
Between different layer growths, the surface crystallinity was checked by means of RHEED.
\textit{Ex-situ} structural characterization was performed using a Bruker D8 Discover high-resolution diffrac\-tometer with monochromatized Cu$_{K, \alpha}$ radiation.
Analysis of the x-ray diffraction and reflectometry data was done utilizing Bruker's DiffracPlus Leptos Software.
For the study of the electronic transport properties, the thin films were patterned by standard UV-photolithography processes followed by low-energy ($\sim$ 500\,eV) Argon-ion etching.
With this, 6-contact Hall bar structures with a bar width of 10\,$\upmu \text{m}$ and an electrode distance of 2000\,$\upmu \text{m}$ were obtained.
Temperature dependent acquisition of the magnetotransport properties was accomplished in a Janis Research cryostat between 5 and 100\,K in magnetic fields up to 7.5\,T.
Additionally, the electrical resistivity was measured in a Scientific Magnetics Helium flow cryostat between 2\,K and 250\,K without field.\\

Hard x-ray Photoelectron Spectroscopy (HAXPES) measurements of the Europium 3d core levels and the valence band at a photon energy of 3.4\,keV were conducted at beamline P22 of the storage ring PETRA III at DESY in Hamburg (Germany) using a time-of-flight momentum microscope to investigate the Eu mean valence \cite{babenkov_high-accuracy_2019, medjanik_progress_2019}.
Samples used for HAXPES were not subjected to the chemical treatment during UV-lithography.

\section{Results and discussion}

\subsection{Structural characterization}

\begin{figure}[htb]
\includegraphics[width=0.48\textwidth]{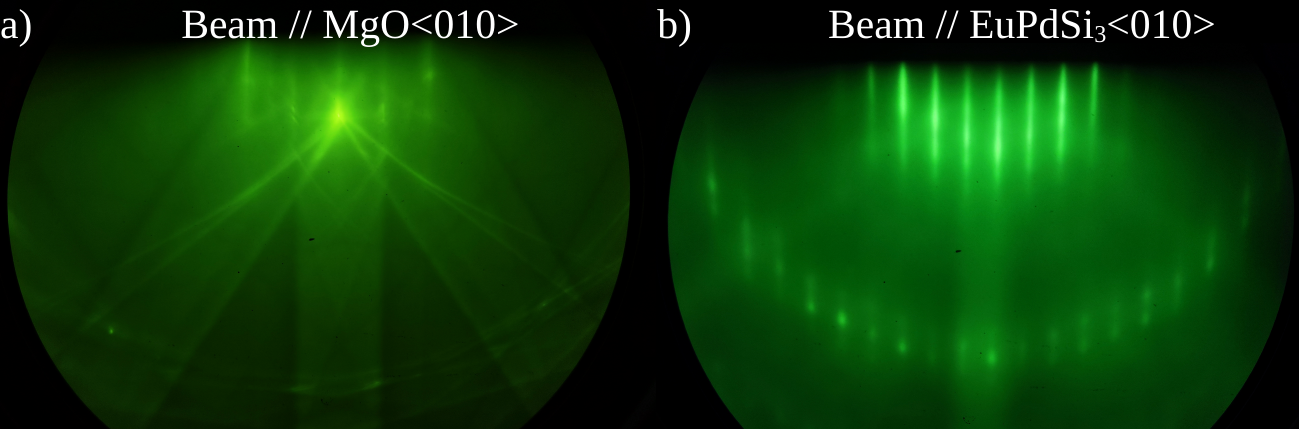}
\caption{\label{rheed} RHEED patterns with electron beam along crystallographic directions as indicated (a) before and (b) after thin film deposition without changing the sample position.}
\end{figure}

Prior to growth, \textit{in-situ} RHEED azimuthal scans show fourfold symmetry upon rotation around the surface normal of the MgO substrate.
After annealing at 1273\,K, the diffraction pattern appears visually sharper, most probably due to the removal of minor surface contaminations. 
The corresponding RHEED image with the electron beam nearly parallel to the MgO$<$010$>$-direction is shown in Fig.\,\ref{rheed}\,a).
Here, sharp diffraction spots appear, which are accompanied by well-defined Kikuchi bands and lines,
pointing to a highly flat and crystalline surface without defects.
After growth, again a fourfold symmetry during sample rotation appears, while the diffraction pattern is mainly composed of narrow streaks.
This points to a crystalline thin film with small mosaicity, where slight misalignments between neighbouring crystallites occur.
Additionally, Kikuchi bands and lines arise, pointing to a laterally well ordered crystalline film.
A comparison of the directions for the main symmetry axes from the substrate and the thin film implies a parallel alignment of their crystallographic \textit{a}-axes. 
In contrast, the RHEED pattern vanishes completely after deposition of the Si capping layer at room temperature.
Thus, an amorphous Silicon overlayer was deposited, without any signs of intermixing between epitaxial film and capping layer.\\

\vspace{-2mm}
\begin{figure}[htb]
\includegraphics[width=0.48\textwidth]{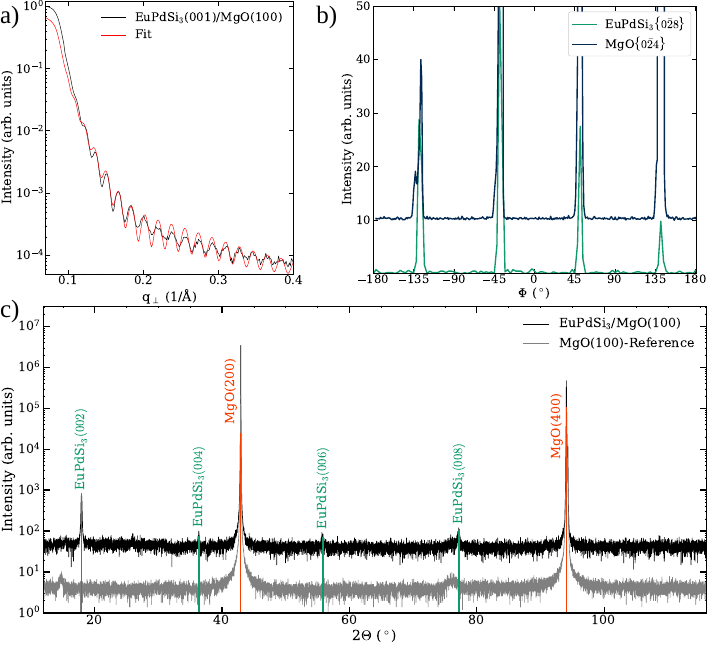}
\caption{\label{xrd} a) X-ray reflectometry scan (black) of a 66\,nm thin film with corresponding fit (red). b) $\phi$-scan around sample normal of MgO$\{0\bar{2}4\}$- and EuPdSi$_3\{0\bar{2}8\}$-reflexes under asymmetric scattering conditions. c) Longitudinal symmetric scan of the same EuPdSi$_3$ thin film (black, upper curve) and a bare MgO(100) substrate as reference (grey, lower curve). The data from the thin film is offset for clarity. The small peak at 78$^\circ$ arises due to scattering from the sample holder.}
\end{figure}

To investigate the structural properties in more detail, \textit{ex-situ} x-ray diffraction (XRD) and reflectometry (XRR) was done under ambient conditions.
The symmetric high angle XRD scan (Fig.\,\ref{xrd}\,c) shows (00$\ell$)-reflexes of EuPdSi$_3$ with even number $\ell$ appearing up to the 8th order, suggesting a well ordered epitaxial thin film. 
Besides these, no other reflexes from secondary phases or other crystallographic directions occur, indicating a high degree of structural order.
Additionally, based on the full width at half maximum (FWHM$_{2\Theta}$\,=\,0.16$^\circ$) of the EuPdSi$_3$(002)-reflex, a mean coherence length of (59\,$\pm$\,3)\,nm can be obtained by fitting (not shown).
The analysis of the positions of the (00$\ell$)-reflexes yields a \textit{c}-axis lattice parameter of \textit{c}\,=\,(9.87\,$\pm$\,0.005)\AA, which is in good agreement with the single crystal value.\\

Besides the symmetric scans, further measurements under asymmetric conditions were performed utilizing $\phi$-scans around the sample surface normal.
Evaluating the precise position of the EuPdSi$_3$(0$\bar{2}$8)-reflex in reciprocal space, a complete relaxation of the in-plane lattice constant with \textit{a}\,=\,4.28\,\AA\, towards the single crystal value is observed.
Furthermore, during the $\phi$-scans, aligning $\omega$ and 2$\Theta$ according to either the position of the EuPdSi$_3$(0$\bar{2}$8)- or the MgO(0$\bar{2}$4)-reflex family, four regularly spaced peaks appear.
As expected for a parallel alignment of the \textit{a}-axes, these reflexes appear under the same $\phi$-angle.
Thus, the $\phi$-scans are in accordance with the epitaxial relationship deduced from RHEED:\\

\noindent MgO\{100\}$\parallel$EuPdSi$_3$\{100\}\,\&\,MgO$<$001$>\parallel$EuPdSi$_3$\!$<$001$>$
\vspace{0.5mm}

It is important to note, that the \textit{a}-lattice parameter of EuPdSi$_3$ (a\,=\,4.28\,\AA) differs by 1\% from that of EuPd$_2$Si$_2$ (a\,=\,4.24\,\AA), while the respective \textit{c}-axes lattice parameters (c\,=\,9.86/9.89\,\AA) differ by only 0.3\%.
Consequently, typical symmetric x-ray diffraction scans along the out-of-plane direction, here parallel to the tetragonal \textit{c}-axes, do not allow an easy identification of the resulting phase.
Interestingly, the EuPdSi$_3$(002)-reflex is more intense than the EuPdSi$_3$(004)-reflex, whereas for EuPd$_2$Si$_2$ the lower-order reflex has a higher intensity, which is in accordance with powder diffraction pattern simulations of the BaNiSn$_3$- and ThCr$_2$Si$_2$-structures.
In consequence, the intensity ratio of the (002)- and (004)-reflexes allows only for a simplified discrimination between both phases.\\

In the small-angle region under symmetric scattering conditions, Kiessig fringes arise as periodic oscillations up to a scattering angle of at least $2\Theta\,\approx\,2.5^\circ$ (corresponding to q$_\perp$\,=\,0.36\,\AA$^{-1}$) due to interference from a smooth interface between the film and substrate (see Fig.\,\ref{xrd}\,a).
Thus, the film thickness can be extracted from the period length of the oscillations using x-ray reflectometry.
As compared to the growth of epitaxial films of EuPd$_2$Si$_2$ \cite{koelsch_clamping_2022}, the maximum angle of occurence of oscillations is reduced, which is indicative of the higher roughness of the EuPdSi$_3$ film.
Fitting a multilayer model, including the Si capping layer, yields best results for a EuPdSi$_3$ film thickness of 66\,nm, a surface roughness of 1.6\,nm and a partially oxidized Si layer with a thickness of approximately 3\,nm using a linear density gradient along the growth direction to account for the formation of SiO$_x$.
The crystalline coherence length $L_c$ in the growth direction is approx. 90\,\% of the total layer thickness, as obtained by XRR, proving a high degree of structural order.  

\subsection{Electronic characterization}

\begin{figure}[htb]
\includegraphics[width=0.48\textwidth]{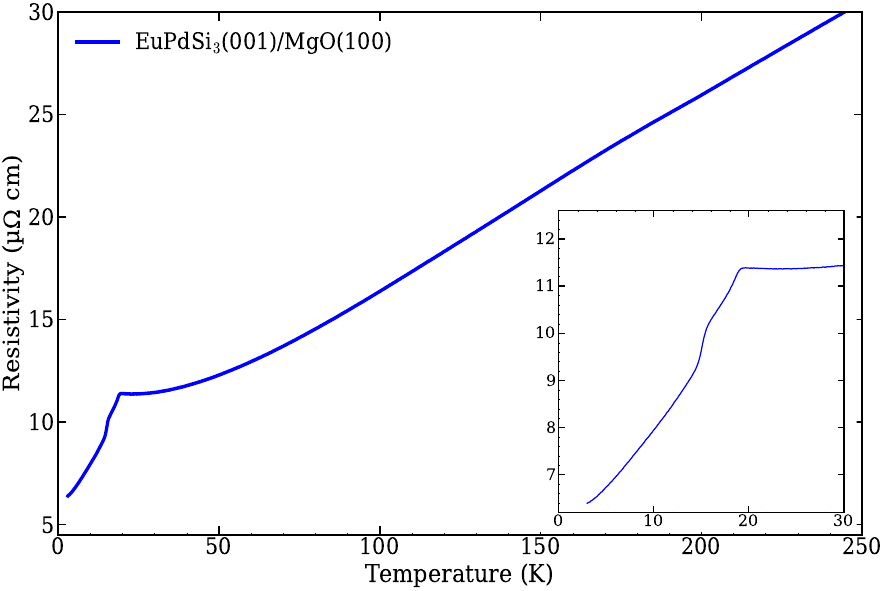}
\caption{\label{transport} Temperature dependence of the electrical resistivity, measured with constant current. The two kinks at low temperature, e.\,g. near 15 and 19\,K, hint at two magnetic transitions in zero applied field.}
\end{figure}

Transport measurements were performed in a cryostat using a constant current of at maximum 1\,mA parallel to the EuPdSi$_3$ \textit{a}-axis, corresponding to a current density of less than 5$\times10^8$\,A/m$^2$.
According to the linear I-V characteristic obtained at different temperatures, heating effects of the structured film can be excluded at these current levels.
Upon cooling, the resistivity decreases monotonically (see Fig.\,\ref{transport}), thus revealing a metallic behavior.
More interestingly, two step-like signatures at approx. T$_{\text{N}1}$\,=\,19\,K and T$_{\text{N}2}$\,=\,15\,K are visible (inset in Fig.\,\ref{transport}), pointing to magnetic phase transitions, resembling bulk behavior \cite{yonehara_single_2020}.
For the resistivity ratio (RR) between 3 and 250\,K, a high value of more than 5 is obtained.
In comparison, epitaxial films \cite{koelsch_clamping_2022} and single crystals \cite{kliemt_strong_2022} of EuPd$_2$Si$_2$ yielded an RR around 1.5-2.\\

\begin{figure}[htb!]
\includegraphics[width=0.48\textwidth]{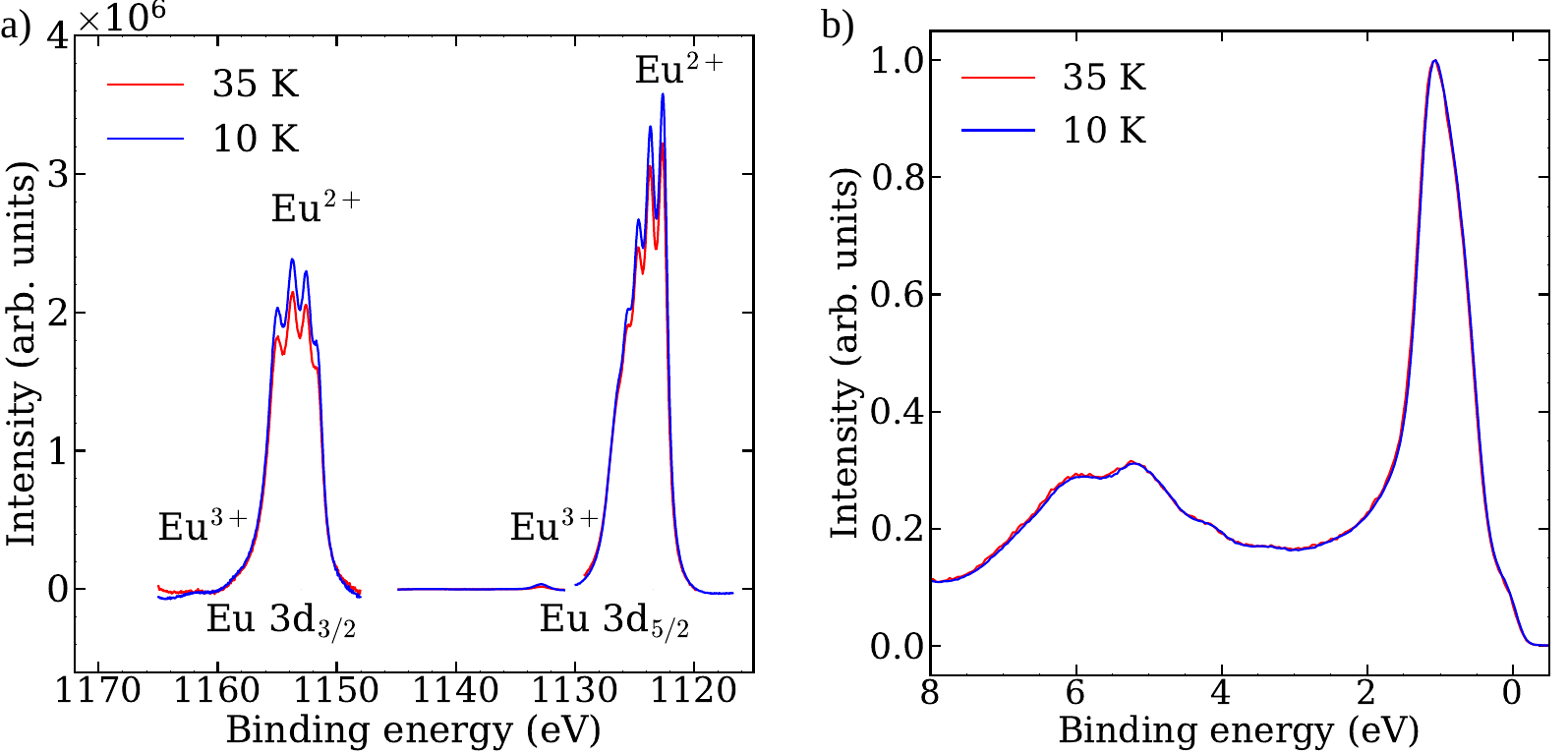}
\caption{\label{haxpes} a) Temperature dependence of the Eu 3d core spectra obtained with a photon energy of h$\nu$\,=\,3.4\,keV at 35\,K (red) and 10\,K (blue). The specified binding energies are referred to the Fermi level E$_F$. b) Corresponding normalized spectra of the valence band region. No significant change occurs upon cooling to 10\,K, reflecting the constant europium valence.}
\end{figure}

Hard x-ray Photoelectron Spectroscopy experiments were performed to study the Eu mean valence well above and below the magnetic phase transitions. 
An incident photon energy of h$\nu$ = 3.4\,keV was selected to obtain more information related to the bulk of the film.
Consequently, it is possible to minimize the influence of a possible surface valence transition, i.\,e. a tendency toward a stable divalent Eu surface, which is not indicative for bulk behavior \cite{martensen_highly_1982, mimura_bulk_2004}.
During the acquisition, the energy resolution is set to about 90\,meV using a Si(111) monochromator and a second post-monochromator crystal.
To determine the mean valence v, the spin-orbit split Eu 3d$_{5/2}$\,-\,Eu 3d$_{3/2}$ spectrum is analyzed, where furthermore a chemical shift separates the Eu$^{2+}$- and Eu$^{3+}$-core levels by approx. 10\,eV for each component.
By analyzing the peak areas of the corresponding electronic configurations, it is possible to derive the mean Europium valence from the spectral weight.
For instance, bulk single crystals of EuPd$_2$Si$_2$ show a strong redistribution of the spectral weight depending on temperature, especially above and below the valence crossover temperature, reflecting directly the valence change behavior \cite{kliemt_strong_2022,fedchenko_valence_2024,fedchenko_electronic_2025}.\\

\begin{figure*}[htpb!]
\includegraphics[width=\textwidth]{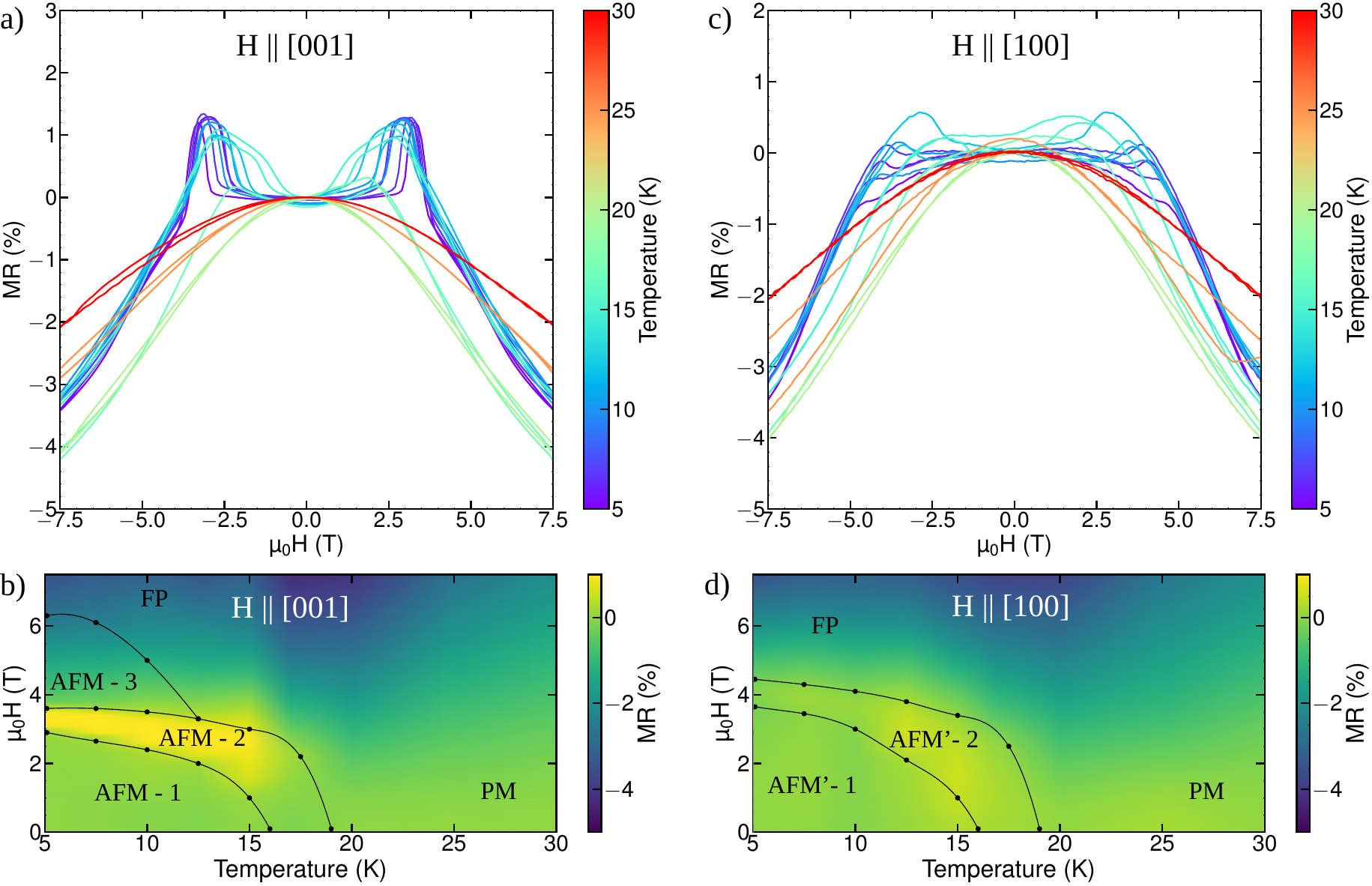}
\caption{\label{cryo-mag} Temperature dependence of the normalized transverse magnetoresistance (MR) measured at constant temperatures with a) magnetic field applied out-of-plane ($\upmu_0$H\,$\parallel$ EuPdSi$_3$[001]) and c) field in-plane ($\upmu_0$H\,$\parallel$ EuPdSi$_3$[100]). b) and d) Corresponding color-coded isothermal MR for both field orientations with increasing field strength. The marked dots indicate the transitions into different antiferromagnetic phases labelled AFM('). FP correspond to the fully field-polarized and PM to the paramagnetic state. The lines are a guide to the eye only.}
\end{figure*}

In the case of epitaxial EuPdSi$_3$ films, nearly no occupation of the Eu$_{3/2}^{3+}$ or Eu$_{5/2}^{3+}$ core levels is detectable, consequently the electronic configuration of Europium is strongly divalent and near v\,=\,$2.00$.
This is remarkable, as in most ternary Eu-based compounds the mean Europium valence is non-integer with values between 2.1 and 2.9 \cite{mimura_bulk_2004, koelsch_clamping_2022, lai_electronic_2022}.
Moreover, there exists no significant change between the high (35\,K) and low (10\,K) temperature spectra with respect to the ratio of the Eu$^{2+}$-/Eu$^{3+}$-components (see Fig.\,\ref{haxpes}).
For the valence band region, the intensity is normalized to the maximum value.
The sharp peak near 1-2\,eV binding energy corresponds to the spectral weight from the Eu$^{2+}$ 4f states.
In consequence, HAXPES measurements using the same photon energy of bulk single crystals of valence-fluctuating EuPd$_2$Si$_2$ show a significant change upon cooling below the valence transition temperature around 160\,K, reflecting the increasing contribution from Eu$^{3+}$ 4f states, see e.\,g. \cite{fedchenko_valence_2024}. 
In contrast, no change of the valence band spectra of the EuPdSi$_3$ thin films is visible upon cooling to 10\,K, proving again a temperature-independent Europium valence.
For this purely divalent Eu-state with 4f$^7$ configuration, the orbital momentum is quenched according to Hund's rule, while the total momentum corresponds to J\,=\,S\,=\,7/2.\\

Returning to the electronic transport measurements under the application of a magnetic field in the out-of-plane direction, a strong change in the magnetoresistance (MR) in a narrow field region is observed below the first magnetic ordering temperature T$_{\text{N}1}$\,=\,19\,K.
Above a temperature dependent saturation field (e.\,g. $\upmu_0$H$_{\text{S}}\approx\,$6.2\,T at 5\,K), the sample is in a fully field-polarized (FP) state, which is in accordance with magnetization measurements on single crystals \cite{yonehara_single_2020, nakashima_magnetic_2026}.
For EuPdSi$_3$ thin films, this regime can be most easily identified by the change in the slope of MR at $\upmu_0$H$_S$, which equals bulk behavior.
At higher temperatures, i.\,e. T\,$\geq$\,20\,K, a simple parabolic shape of MR(H) with negative curvature appears, pointing to the paramagnetic state.
Additionally, abrupt changes below T$_{\text{N}1}$ in the slope or shape of the MR at fields below full saturation are characteristic signatures for phase transitions of the respective magnetic order, enabling the mapping of different magnetic phases, as discussed below.\\

Most prominently, the magnetoresistance shows a pronounced hysteresis effect for T\,$<\,$T$_{\text{N}2}$\,$\approx$\,15\,K at intermediate fields between 2\,T\,\,$\lesssim$\,\,$\upmu_0$H\,\,$\lesssim$\,\,3\,T during a complete measurement cycle (see Fig.\,\ref{cryo-mag}\,a) for the difference between increasing and decreasing magnetic field.
Furthermore, a close inspection of the low temperature data (T\,$\lesssim$\,12.5\,K) reveals another transition at higher fields. 
At temperatures above 13\,K no signatures for this phase are visible in the MR measurements.
Additionally, measurements of the Hall effect (HE) show a linear contribution above $\upmu_0$H$_{\text{S}}$ according to the normal HE (NHE).
More interestingly, at lower fields the contribution of the anomalous Hall effect (AHE) (proportional to the magnetization in collinear ferromagnets) is an order of magnitude larger than that of the NHE, which reflects the non-collinear spin arrangement (not shown).
The resulting H-T phase diagram in Fig.\,\ref{cryo-mag}\,c) with H\,$\parallel$\,[001] summarizes all the transitions into the three different antiferromagnetic structures (abbreviated as AFM-1 to -3), while the displayed data reflects the measurements with increasing field strength.\\

For applied field parallel to EuPdSi$_3$[100] (in-plane), with current flow along EuPdSi$_3$[010], the behavior of the MR is similar.
Again, for T\,$\geq$\,20\,K a simple parabolic shape of MR appears.
More interestingly, hysteresis effects in the AFM'-1 phase, appearing below 15\,K, expand down to zero applied field, which is in contrast to the out-of-plane field configuration (see Fig.\,\ref{cryo-mag}\,b).
However, the change in the shape of the MR is less pronounced, and the maximum is shifted to higher fields, e.\,g., around 4\,T for T\,$\leq$\,15\,K.
In the single crystal case, a similar shape is observed, although the maximum appears at even higher transition fields, e.\,g., around 6\,T at 5\,K \cite{ocker_eupdsi3_2026}.
According to the magnetization and susceptibility measurements on the bulk material \cite{ocker_eupdsi3_2026}, upon application of a magnetic field along either [100] or [001], several metamagnetic transitions can be tracked by resistivity measurements equally well.
In particular, near zero field, a complex antiferromagnetic type of order can be deduced, where some sort of spin canting will arise upon increasing the external field. 
Comparing the results from both field configurations, the EuPdSi$_3$[001] direction corresponds to the easy axis, which is also the case for the isostructural EuPtSi$_3$ compound \cite{bauer_magnetic_2022}.

\section{Conclusion}
Using molecular beam epitaxy, we demonstrate epitaxial film growth with high-quality of tetragonal EuPdSi$_3$ on annealed MgO(100) substrates.
A simple cuboid-on-cube growth mode with the epitaxial relationship $\text{MgO}\{100\}$\,$\parallel$\,EuPdSi$_3\{100\}$ and MgO$<$001$>$\,$\parallel$EuPdSi$_3$\,$<$001$>$ is established by means of \textit{in-situ} electron and \textit{ex-situ} x-ray diffraction.
Since the \textit{a}- and \textit{c}-axis lattice constants are equal to their bulk crystal values at room temperature, a fully relaxed growth occurs.
This is remarkable, as the \textit{a}-axis lattice constants between MgO and EuPdSi$_3$ differ by nearly 1\% and suggests a thin interfacial layer, in which strain relaxation occurs.
Furthermore, the results regarding the growth behavior are in accordance with the epitaxial growth of EuPd$_2$Si$_2$ films on MgO(100), having a comparable misfit ($\sim$0.5\%), see \cite{koelsch_clamping_2022}.\\

The grown films were studied by means of electronic transport and Hard x-ray Photoelelectron Spectroscopy, clarifying the electronic configuration of Europium and examining the magnetic transitions.
Temperature dependent HAXPES measurements reveal a constant Europium valence of 2.0, giving opportunity for magnetic ordering with a high magnetic moment of approximately 7\,$\upmu_{\text{B}}$/Eu atom, according to Hund's rule.
In particular, no change of the Eu valence is observed down to 10\,K.
In consequence, the two phase transitions at zero applied magnetic field can be accounted solely to different types of magnetic ordering upon cooling.\\

Under the application of an external magnetic field along either the [100]- or [001]-direction, differently ordered phases are observed in electronic transport measurements.
The corresponding H-T phase diagrams for the two field orientations are in good accordance with single crystal data \cite{nakashima_magnetic_2026, ocker_eupdsi3_2026}, while slight differences point to the influence of thermally induced strain through the substrate in the thin film case.
More interestingly, for the out-of-plane field configuration a large anomalous Hall effect was found.\\

Due to a strong coupling between the in-plane lattice constants in the related EuPd$_2$Si$_2$/MgO system, the valence transition, known from EuPd$_2$Si$_2$ single crystals, is completely suppressed, instead, magnetic ordering takes place below T$^*\approx$\,19\,K \cite{koelsch_clamping_2022}.
Here, similar effects are to be expected with respect to the \textit{a}-lattice constant of EuPdSi$_3$ on MgO substrates at low temperatures, i.\,e. a significant deviation from the unconstrained behavior in EuPdSi$_3$ single crystals.
In consequence, this mechanical biaxial in-plane strain influences the magnetic structures, whereas no significant change in the ordering temperatures at zero applied field is observed.
At this time, we can only speculate on the concrete magnetic structures and the exact influence of this thermally induced strain.
In particular, a comparison with similar, isostructural compounds yields differently modulated antiferromagnetic phases.
The transition at lower temperature might indicate the locking from an incommensurate to a commensurate mode, see e.\,g. \cite{maurya_enhanced_2015}.
To resolve the magnetic structure in detail, resonant elastic x-ray scattering (REXS) at the Eu L$_{\text{II}}$ absorption edge (7.61\,keV) is required.
As was shown recently for single crystals of isostructural EuPtSi$_3$ with comparable ordering temperatures of 16\,K and 17\,K, tuning the incident x-ray energy to the absorption edge significantly enhances the contribution of magnetic ordering to the scattering signal \cite{simeth_resonant_2023}.
Research along these lines is currently under way.

\begin{acknowledgments}
Funded by the Deutsche Forschungsgemeinschaft (DFG, German Research Foundation) - TRR 288 - 422213477 (projects A04, B02 and B04).
The funding of the instrument by the Federal Ministry of Education and Research (BMBF) under the framework program ErUM is gratefully acknowledged. 
We thank DESY (Hamburg, Germany), a member of the Helmholtz Association HGF, for the provision of experimental facilities. 
Parts of this research were performed at PETRA III using beamline P22.
\end{acknowledgments}

\bibliographystyle{unsrtnat}

\end{document}